\documentclass[twocolumn,showpacs,prl,floatfix,superscriptaddress]{revtex4}

\usepackage{color} 
\usepackage{tabularx} 
\usepackage{epsfig}
\usepackage{amsmath} 
\usepackage{amssymb} 
\usepackage{graphicx}
\usepackage{wasysym}
\usepackage{times}

\input cyracc.def
\newfam\cyrfam
\font\tencyr=wncyr10
\font\sevencyr=wncyr7
\font\fivecyr=wncyr5
\def\cyr{\fam\cyrfam\tencyr\cyracc}
\textfont\cyrfam=\tencyr \scriptfont\cyrfam=\sevencyr
\scriptscriptfont\cyrfam=\fivecyr
\usepackage{makerobust}
\MakeRobustCommand{\cyr}

\begin{document}

\title{Simplest model to study reentrance in physical systems}

\author{Creighton K.~Thomas}
\affiliation {Department of Physics and Astronomy, Texas A\&M University,
College Station, Texas 77843-4242, USA}

\author{Helmut G.~Katzgraber} 
\affiliation {Department of Physics and Astronomy, Texas A\&M University,
College Station, Texas 77843-4242, USA}
\affiliation {Theoretische Physik, ETH Zurich, CH-8093 Zurich, Switzerland}

\date{\today}

\begin{abstract}

We numerically investigate the necessary ingredients for reentrant
behavior in the phase diagram of physical systems. Studies on the
possibly simplest model that exhibits reentrance, the two-dimensional
random bond Ising model, show that reentrant behavior is generic
whenever frustration is present in the model.  For both discrete and
continuous disorder distributions, the phase diagram in the
disorder--temperature plane is found to be reentrant, where for some
disorder strengths a paramagnetic phase exists at both high and low
temperatures, but an ordered ferromagnetic phase exists for
intermediate temperatures.

\end{abstract}

\pacs{75.10.Hk, 05.70.Fh, 64.60.Cn, 64.60.an}

\maketitle

Reentrance in a thermodynamic phase diagram presents a counterintuitive
scenario: one phase exists inside some closed temperature range,
with transitions to the {\em same} second phase at both higher and
lower temperatures.  If one of these phases is more ordered than
the other one, then one of the phase transitions must violate the
intuition that lower temperature phases are more ordered than higher
temperature phases.  In the context of solid-liquid transitions,
this is known as ``inverse melting'' or ``inverse freezing.''
A variety of materials have been shown to have reentrant phase
diagrams.  For example, Rochelle salt was found to be
a ferroelectric with two Curie points; the ordered ferroelectric
phase occurs only between these two temperatures \cite{jona:62}.
More recently, similar phase diagrams have been seen for superconducting 
vortices \cite{fertig:77-ea,avraham:01-ea}, liquid crystals
\cite{cladis:75}, miscibility in solutions \cite{zaitsev:86-ea},
polymeric materials \cite{vanruth:04}, ferromagnetism in
semiconductors \cite{krivoruchko:10-ea}, denaturation of DNA \cite{hanke:08-ea},
and many other systems \cite{cladis:88,schupper:05}.

Theoretically, a number of model systems have been found with
reentrant phase diagrams. The Ising model on a Union Jack lattice with
frustrated anisotropic interactions was shown to be reentrant in a
narrow region of parameter space \cite{vaks:66-ea,morita:85}.  In the
fully-frustrated Villain model \cite{villain:77} the ground state
of the system is seen to be disordered, while the low-lying excited
states favor ferromagnetic ordering; in this case the ordering is due
to the emergence of ferrimagnetism \cite{villain:80-ea}.  Reentrant
ferromagnetism has also been seen in models of semiconductors because
the carrier density increases with temperature \cite{petukhov:07-ea};
in models of solid hydrogen, reentrance is due to quantum
fluctuations \cite{hetenyi:99-ea}.
Most prominently, frustrated spin-glass models have been shown to
include the complexity necessary to describe rather generic reentrant
scenarios
\cite{berker:81,schupper:04,crisanti:05,paoluzzi:10-ea,ferreira:10-ea},
albeit for models with complex Hamiltonians.

Here we show numerically that the two-dimensional random-bond Ising
model (RBIM)---possibly the simplest disordered spin model with
frustration---generically possesses reentrance in its phase diagram
for {\em both} discrete and continuous disorder distributions.
It is given by the Hamiltonian
\begin{equation}
\textstyle{
\mathcal{H} = -\sum_{\langle i j \rangle} J_{ij} s_i s_j 
}
\label{eq:H}
\end{equation}
with an $L\times L$ square toroidal grid \cite{comment:bc} of
Ising spins $\{s_i \in \pm 1\}$ and quenched nearest-neighbor
random couplings
$J_{ij}$. These random couplings
are most commonly chosen from either a bimodal ($\pm J$) or Gaussian
distribution.  We emphasize that this Hamiltonian is much simpler than
the disordered spin models typically used to study inverse freezing
(e.g., the ``simplest'' model of inverse freezing \cite{crisanti:05}).
Furthermore, the phase diagram of this model is uncomplicated.
At any finite temperature, with a disorder distribution biased toward
ferromagnetic interactions, only ferromagnetic and paramagnetic phases
are possible \cite{fisher:88}.  Because reentrance scenarios often occur
in complex phase diagrams with many distinct phases, such
a clean example is difficult to find in reentrant models.

\paragraph*{The random-bond Ising model.---}
\label{sec:eaisg}

The RBIM is widely studied as a standard model of disordered systems;
it is one of the simplest models which includes the disorder and
frustration needed to exhibit a complex glassy behavior at low
temperatures \cite{binder:86}. These same elements, especially
frustration, are necessary for the reentrant behavior in the phase
diagram seen here. Furthermore, the model and cousins on more complex
lattice geometries are of paramount importance across disciplines
and more recently have found widespread use in the computation of
the error stability of topologically-protected quantum computing
proposals \cite{kitaev:03,dennis:02-ea,katzgraber:09c-ea}.  For quantum
computing applications, the order-disorder transition corresponds to
the maximum error rate at which quantum operations may be performed
with high fidelity.  This relation is particularly apparent because
the quantity we use to identify the phase transition in the RBIM is
closely related to the error rate in the quantum computing proposals.

We study the phase diagram for this model with both discrete
and continuous bond disorder
$\{J_{ij}\}$.
For the discrete $\pm J$ distribution, bond values are chosen
according to
\begin{equation}
P(J_{ij}) = p\delta(J_{ij}-J) + (1-p)\delta(J_{ij}+J). 
\label{eq:pmJdistr}
\end{equation}
The pure Ising model is recovered for $p \to 1$, while the
strong-disorder case is typically studied for $p=0.5$.  The disorder
strength $q=1-p$ gives the deviation from a pure ferromagnet, but note
that the free energy $F$ of the model is symmetric under reflection
about $p=0.5$, so $F(T,p)=F(T,q)$.  Nishimori has shown that, due
to extra symmetries of the problem, a number of quantities, such
as the internal energy of the system, may be computed exactly when
the equality $(1-p)/p = \exp (-2 J / T )$ holds \cite{nishimori:81}.
This equality is called the Nishimori line. The location of the phase
transition on the Nishimori line, $(T_c^*,p_c^*)$, has been identified
with a multicritical point \cite{ledoussal:88}.  Of interest here is
that the magnetization for a given value of $p$ must be greatest on the
Nishimori line, so that the ferromagnetic phase must not exist at any
temperature for $p < p_c^*$ \cite{nishimori:81}. This implies that the
phase diagram below the Nishimori line is either vertical or reentrant.
It has been argued analytically that the vertical case is the correct
one \cite{kitatani:92,ozeki:93}, although numerical studies
with $\pm J$ disorder suggest a reentrant 
phase diagram \cite{nobre:01,wang:03-ea,amoruso:04,parisen:09-ea}.

For the Gaussian distribution, bonds are chosen from
\begin{equation}
P(J_{ij}) = (2\pi\widetilde{J}^2)^{-1/2}
	     \exp\left[-(J_{ij}-J)^2/2\widetilde{J}^2\right].
\label{eq:gaussint}
\end{equation}
The pure Ising model is recovered for $\widetilde{J}\to
0$, while the strong disorder case occurs when
$\widetilde{J}=1$ and $J=0$.  The free energy of the model is
symmetric under reflection about $\widetilde{J}=0$.  It is customary
\cite{mcmillan:84,melchert:09} to define a disorder strength parameter
$r = \widetilde{J} / J$. In the case of Gaussian disorder, the Nishimori
line \cite{nishimori:81} is given by
$\widetilde{J}^2/T = J$.
Here, the multicritical point $(T_c^*,r_c^*)$ is also the largest
value of $r$ for which a ferromagnetic phase may exist, so that the
phase diagram for Gaussian disorder is also expected to be either
vertical or reentrant.

Unfortunately, the exact location of the multicritical point has
not been calculated for either disorder distribution; numerical
estimates for $\pm J$ disorder have given values of $q_c^*=0.1094(2)$
\cite{honecker:01-ea}, $0.1093(2)$ \cite{merz:02}, and $0.10917(3)$
\cite{parisen:09-ea}, while estimates for Gaussian disorder include
$r_c^*=0.97945(4)$ \cite{picco:06-ea} and $0.9811(3)$, where the latter number was
quoted as $1/r_c^*=1.0193(3)$ \cite{queiroz:09}.
Exact efficient ground state algorithms exist for
studying this model at zero temperature \cite{barahona:82}, although
the use of such techniques is complicated because they typically do
not handle degeneracy well.  Nevertheless, the values obtained are
in the vicinity of $q_c(T=0) = 0.103$ \cite{wang:03-ea,amoruso:04} and
$r_c(T=0)=0.970(2)$ \cite{melchert:09},
and clearly are not consistent with the values at the multicritical point.

One could still imagine a scenario where the $T=0$ behavior is
significantly different from all finite-temperature behaviors: this
is the case for the strong-disorder region of the phase diagram, where
the $T=0$ spin-glass ``phase'' does not exist at nonzero temperatures
\cite{fisher:88}.  Low-but-finite-temperature simulations are typically
quite difficult due to long equilibration times, and the difference
between the numerical estimates is quite small, so few points in
between have been probed.  The numerical measurement of the location
of the phase transition at finite temperatures below the Nishimori
line \cite{merz:02,parisen:09-ea} has required thorough finite-size
scaling extrapolation. Nevertheless, computations of $p_c(T)$ for
two points below the Nishimori line have shown intermediate results
\cite{parisen:09-ea}. Here, our more efficient technique allows us to
see the reentrance far more clearly than in previous studies.

\paragraph*{Simulation details.---}
\label{sec:sim}

To investigate the disorder-temperature phase diagram, we numerically identify
the parameters that give the ferromagnet-paramagnet phase transition using a
Pfaffian technique. Because this phase transition is of second order, the
magnetization, which is the natural order parameter, is continuous. Quantities
such as the Binder ratio \cite{binder:81} have been developed to most precisely
determine the location of a phase transition. Here, we use a different
quantity, which is more easily computed when the partition function is directly
available.

We start be defining an extended Hamiltonian
(see Ref.~\cite{thomas:07}) where the boundary conditions
are allowed to vary, with both periodic and antiperiodic cases
being allowed.  The extended Hamiltonian is given by
$\mathcal{H}^* = -\sum_{\langle ij \rangle}J_{ij} s_i s_j \sigma_{ij}$,
with $\sigma_{ij}=1$ except on one vertical column of horizontal bonds
where $\sigma_{ij}=\sigma_v$ and one horizontal row of vertical bonds,
where $\sigma_{ij}=\sigma_h$; the $\sigma_{h,v}=\pm 1$.  Defining a
configuration $\{\{s_i\},\sigma_v,\sigma_h\}$ of the system now
requires specifying the spin values as well as the boundary conditions.
The partition function,
$\mathcal{Z} = \sum_{\{s_i=\pm 1\}}\exp(-\mathcal{H}/T)$,
may also be extended to
$\mathcal{Z^*} =
\sum_{\{s_i=\pm1\},\sigma_v=\pm1,\sigma_h=\pm1}\exp(-\mathcal{H^*}/T)$.
Using a Pfaffian technique, $\mathcal{Z^*}$ may be directly evaluated
from the same computation that produces $\mathcal{Z}$ with {\em no
additional computational effort}.  In this extended system we can compute
the probability ${\cyr zh}$ that the boundary conditions are periodic
in both directions, given by
\begin{equation}
{\cyr zh } = \mathcal{Z}/\mathcal{Z^*}.
\end{equation}
When the boundary conditions in a direction are changed from periodic
to antiperiodic a system-spanning domain wall of length scale
$L$ is imposed across the system.  In the ferromagnetic phase,
a system-spanning domain wall is energetically unfavorable: the
periodic-periodic case, $\sigma_v=\sigma_h=1$, has lower free energy
by an amount proportional to $L$ and therefore ${\cyr zh} = 1$ as
$L\to \infty$.  In the paramagnetic phase, the boundary conditions
do not significantly influence the thermodynamics of the system,
and all four cases have equal weight, so that ${\cyr zh} = 1/4$
as $L\to \infty$.  At the phase transition, ${\cyr zh}$ approaches
a constant {\em independent} of $L$ (up to finite-size corrections), 
i.e., ${\cyr zh} \sim \tilde{Z}[L^{1/\nu}(T - T_c)]$.
This measure is somewhat more sensitive than the free energy of a
system-spanning domain wall because it allows for system-spanning
domain walls in either or both directions.
Note also that $1-{\cyr zh}$ is equivalent to the error rate
discussed in Ref.~\cite{wang:03-ea}. Thus the method we
introduce here can be applied {\em generically} to the study of
topologically-protected quantum computing proposals.

We have computed ${\cyr zh}$ to investigate the phase diagram of the
random-bond Ising model using a Pfaffian technique.  With periodic
boundary conditions, one may exactly compute $\mathcal{Z}$ by summing
the Pfaffians of four related Kasteleyn matrices \cite{thomas:09}.
The partition function of a system of $N=L\times L$ spins may be
computed in $\mathcal{O}(N^{3/2})$ operations, allowing for exact
partition function evaluation in systems up to $L=512$.  By choosing
different signs for the terms in the sum, it is also possible, without
computing any additional Pfaffians, to compute $\mathcal{Z}$ for {\em
all four} boundary conditions \cite{thomas:09}.  Thus by computing
$\mathcal{Z}$, we obtain $\mathcal{Z^*}$, and therefore ${\cyr zh}$
for no extra effort.

\paragraph*{Results.---}
\label{sec:res}

\begin{figure}[tb]
\includegraphics[width=\columnwidth]{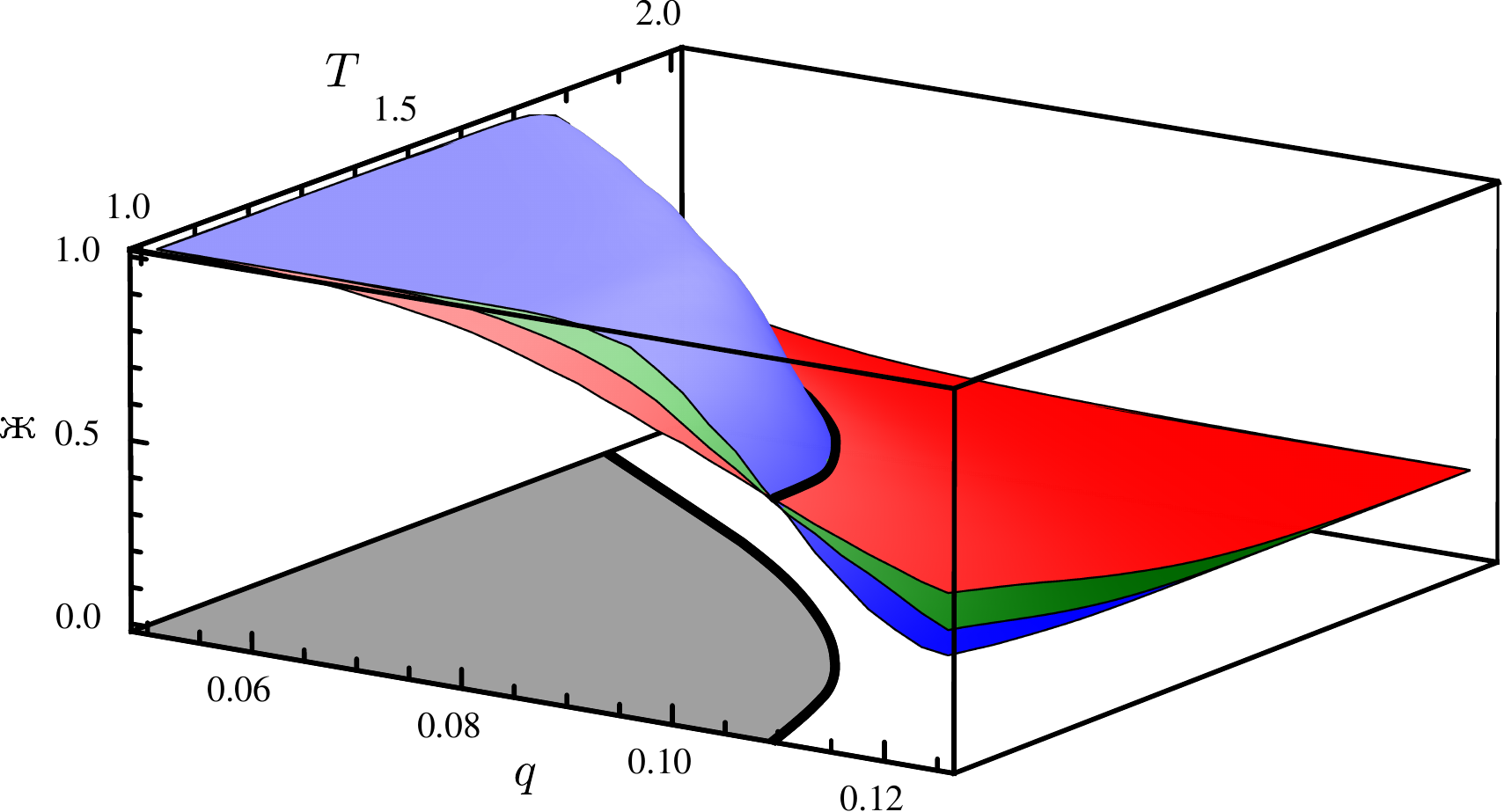}
\caption{(Color online) 
Order probability ${\cyr zh}$ as a function of temperature $T$ and
disorder strength $q$ for the RBIM with $\pm J$
interactions for system sizes $L = 16$, $32$, and $64$ (shallowest to
steepest). Finite-size scaling corrections are small, 
i.e., the surfaces cross cleanly at 
the phase boundary (shown in the projection onto
the plane). Error bars 
are smaller than the thickness of the surfaces.
}
\label{fig:surface}
\end{figure}

\begin{figure}[tb]
\includegraphics[width=0.95\columnwidth]{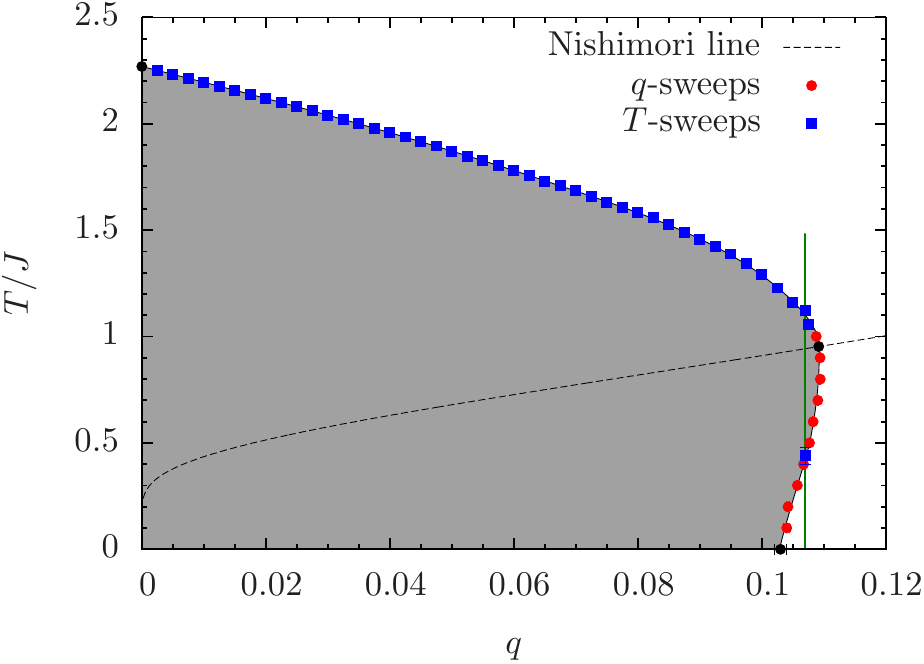}
\caption{(Color online)
Phase diagram of the two-dimensional random-bond Ising model with
$\pm J$ interactions [Eq.~(\ref{eq:pmJdistr})].  The 
shaded region is ferromagnetic, while the white region is paramagnetic.
The boundary shown between these two regions is a guide to the eye.
Some phase boundary points are computed
by doing a scaling collapse varying $T$ (squares), while others
are from a scaling collapse varying $q$ (circles).  Statistical error
bars which are smaller than the symbol size are not visible.  For $q < 0.05$,
we use the results from Ref.~\cite{ohzeki:11-ea}, which used a
related technique. The black circle for $q=0$ is the exact result
for the pure Ising model. The point on the Nishimori line is from
Ref.~\cite{parisen:09-ea} and the $T=0$ point where the boundary touches
the axis is from Ref.~\cite{amoruso:04}.
}
\label{fig:pmJ_phase_diagram}
\end{figure}

\begin{figure}[tb]
\includegraphics[width=0.95\columnwidth]{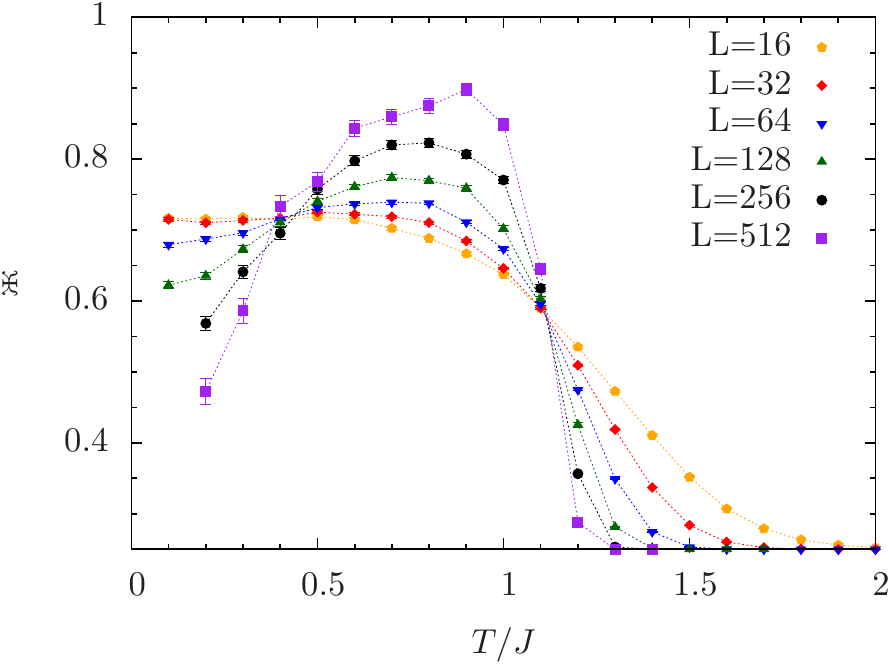}
\caption{(Color online)
Order probability $\cyr zh$ as a function of temperature for $q = 0.107$
(vertical line in Fig.~\ref{fig:pmJ_phase_diagram}) and $\pm J$
disorder. The data show two crossings, illustrating the
existence of two transitions and therefore reentrance in the phase
diagram.
}
\label{fig:pmJslice}
\end{figure}

Figure \ref{fig:surface} shows the order probability ${\cyr zh}$ as
a function of temperature $T$ and disorder strength $q$ for the
random-bond Ising model with $\pm J$ interactions and different
system sizes. The data do cross at a line (see projection) that
corresponds to the phase boundary, thus illustrating that the
approach used works well.  To extract the best estimate of the critical
temperature $T_c$ for a given value of $q$ we perform a finite-size
scaling of the data with $T_c$ and $\nu$ as free parameters. After
performing a Levenberg-Marquard minimization of the chi$^2$ of the
best fit to a third-order polynomial we estimate statistical errorbars
by wrapping the process in a bootstrap analysis.  The phase diagram
for $\pm J$ disorder is shown in Fig.~\ref{fig:pmJ_phase_diagram}
and clearly shows reentrant behavior.  We have these scaling collapses 
for data at fixed $q$ (squares) and $T$ (circles). To further highlight
the reentrant behavior, in Fig.~\ref{fig:pmJslice} we show the
order probability as a function of temperature $T$ for $q = 0.107$
(vertical line in Fig.~\ref{fig:pmJ_phase_diagram}). The data
show two crossings, therefore clearly indicating that the phase 
diagram is paramagnet--ferromagnet--paramagnet with two distinct
transitions.

\begin{figure}[tb]
\includegraphics[width=0.95\columnwidth]{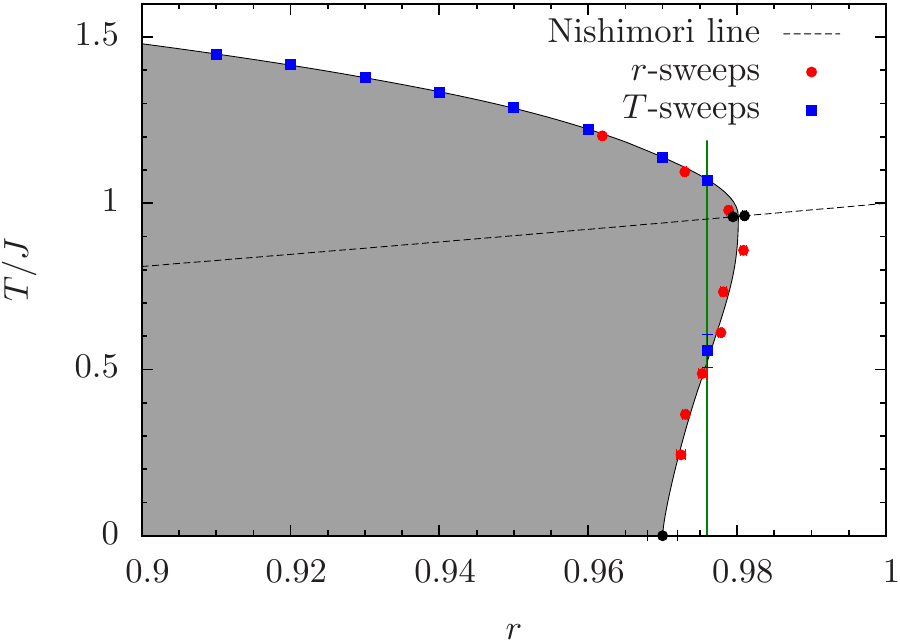}
\caption{(Color online)
Phase diagram of the two-dimensional random-bond Ising model with
Gaussian disorder [Eq.~(\ref{eq:gaussint})].  The shaded
region is ferromagnetic, while the white region is paramagnetic.
The black points on the Nishimori line are from Refs.~\cite{picco:06-ea} (left) and
\cite{queiroz:09} (right), and the point at $T=0$ is from
Ref.~\cite{melchert:09}.
The boundary shown between these two regions is a guide to the eye.
}
\label{fig:gaussian_phase_diagram}
\end{figure}

\begin{figure}[tb]
\includegraphics[width=0.95\columnwidth]{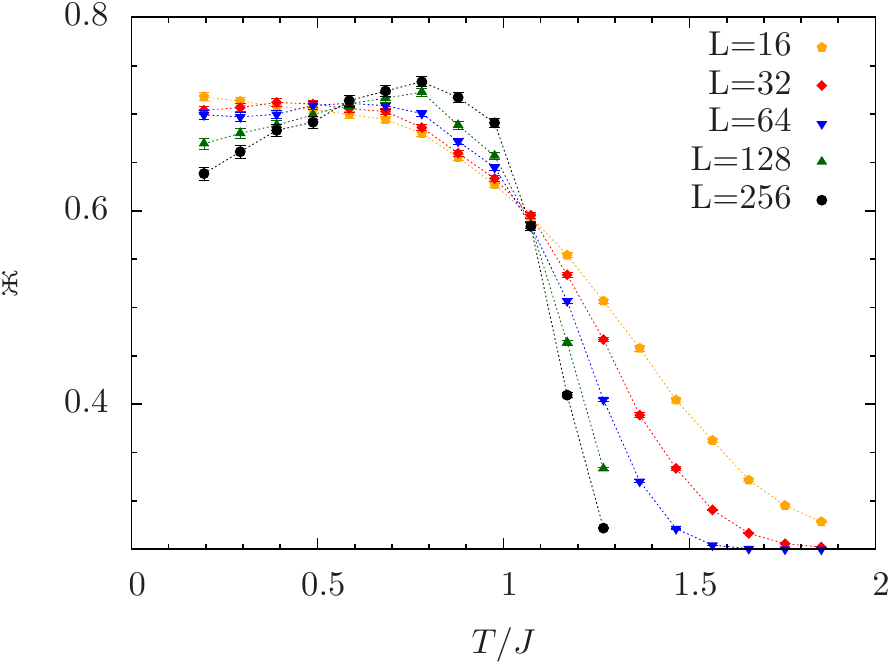}
\caption{(Color online)
Order probability $\cyr zh$ as a function of temperature for $r = 0.976$
(vertical line in Fig.~\ref{fig:gaussian_phase_diagram}) and Gaussian
disorder. Although the approach to the thermodynamic limit is slow, the data
clearly show two crossings, illustrating the
existence of two transitions and therefore reentrance in the phase
diagram.
}
\label{fig:gaussJslice}
\end{figure}

Finally, in Figs.~\ref{fig:gaussian_phase_diagram} and
\ref{fig:gaussJslice} we show data for Gaussian disorder.  Reentrance
is clearly present, albeit much weaker than for the $\pm J$ case:
the ratio $r_c(T=0)/r_c^*\approx0.99$ is much closer to $1$ than
$q_c(T=0)/q_c^*\approx0.94$ for the $\pm J$ case.  These results
show clearly that reentrance is a generic feature of this model
when disorder and frustration are present. The fact that the case
with Gaussian disorder has a much weaker effect suggests that the
ground-state entropy might play a role but is not strictly necessary.
In fact, studying a model where a continuous transition between the
$\pm J$ and Gaussian cases can be tuned \cite{pelikan:04-ea} might
help in elucidating this behavior, but it would be computationally
extremely expensive.  This tuning could change the magnitude of the
reentrance, but it appears that the phase transitions are in the
same universality class: the critical exponent $\nu$ is consistent
with $\nu\approx1.5$ for all points below the Nishimori line.
For $\pm J$ disorder, we find an aggregate $\nu=1.49(4)$ while for
Gaussian disorder, $\nu=1.52(5)$, in line with previous studies
\cite{wang:03-ea,amoruso:04,merz:02,parisen:09-ea,melchert:09,picco:06-ea,queiroz:09}.

\paragraph*{Summary and Discussion.---}
\label{sec:sum}

We have shown that the random-bond Ising model in two dimensions---a
simply posed model with only two phases---generically possesses
reentrance in its phase diagram.  The disorder and frustration
present in this model are responsible for this counterintuitive
result.  This is likely related to the ``order by disorder''
seen in the Villain fully frustrated model \cite{villain:80-ea}.
In the fully-frustrated case, ferromagnetic strips are completely
decoupled from one another in the ground state, but the low-lying
excitations have a weak ferromagnetic interaction among the strips.
In the RBIM, the ground state consists of ferromagnetic domains of
a size and energy scale set by the disorder strength.  When these
are dense enough to percolate throughout the system, ferromagnetic
ordering will cease at $T=0$.  However, for a range of parameters these
domains can be coupled strongly enough in the low-lying excitations
to produce the reentrant behavior in this model.  It should be noted
that the reentrance scenario shown here is particular to two space
dimensions; the random bond Ising model in higher dimensions has a
low-temperature spin glass phase so that if the ferromagnetic phase
only exists for intermediate temperatures, the low-temperature phase
would be a spin-glass phase and not the same paramagnetic phase found
at high temperatures.

\begin{acknowledgments} 

H.G.K.~acknowledges support from the SNF (Grant No.~PP002-114713).
The authors acknowledge ETH Zurich for CPU time on the Brutus cluster.

\end{acknowledgments}

\bibliography{refs,comments}

\begin{thebibliography}{47}
\bibitem[{com()}]{comment:bc}
\bibinfo{note}{The toroidal geometry ensures periodic boundary conditions and
  hence minimizes corrections to scaling.}
\end{thebibliography}

\end{document}